\documentstyle[graphicx,pifont,amssymb,amsmath]{mn}

\title[Chandra HRC and HST observations of NGC6240]
{Chandra HRC and HST observations of NGC6240: resolving the AGN and Starburst}

\author[Lira etal.]
{P.~Lira,$^{1,2}$ M.J.~Ward,$^1$ A.~Zezas,$^3$ S.S.~Murray$^3$\\
$^1$ Department of Physics \& Astronomy, University of Leicester, Leicester LE1 7RH, UK\\
$^2$ Departamento de Astronom\'{\i}a, Universidad de Chile, Casilla 36-D, Santiago, Chile\\
$^3$ Harvard-Smithsonian Center for Astrophysics, 60 Garden St., Cambridge MA, 02138, USA\\} 

\begin{document}

\maketitle

\begin{abstract} 

We present high spatial resolution X-ray Chandra HRC and HST WFPC2
H$\alpha$ observations of the prototypical infrared luminous galaxy
NGC6240. The central region of this system shows a remarkably complex
morphology, with filaments and loops observed in the optical and
X-rays. The total X-ray luminosity is dominated by the extended
emission. Both nuclei are clearly detected in the HRC image and both
appear to be extended. The energetics of the nuclei imply that the
southern nucleus is the more plausible counterpart to the obscured
AGN. The overall SED of the galaxy is in good agreement with a blend
of starburst and AGN components which have similar bolometric
luminosities, $L_{bol}\sim5\times10^{45}$ ergs s$^{-1}$, with the
starburst dominating the observed continuum in the near-IR (K-band),
optical and soft X-ray bands.

\end{abstract}


\section{Introduction}

NGC6240 is a prototypical luminous infrared galaxy. Its high IR
luminosity ($L_{\rm IR} \la 10^{12} L_{\sun}$) and proximity ($d =
147$ Mpc, for $z = 0.0245$, H$_\circ=50$ km s$^{-1}$ Mpc$^{-1}$) have
made it the subject of numerous studies at all wavelengths. Its
central region exhibits an extremely complex optical morphology with
two distinctive nuclei, whilst at larger radii a distorted disc-like
system with extended tidal tails is observed. ROSAT HRI observations
of the galaxy show that extended luminous emission can be traced out
to a radius of $\sim 20\arcsec$ (Komossa, Schulz \& Greiner, 1998).

Starburst activity dominates the emission of NGC6240 at most
wavelengths. Optical spectroscopy classifies the nuclei as LINER
(Veilleux et al. 1995), whilst integral field near-IR observations
show that the K band is dominated by red supergiants (Tecza et al.
2000).  Mid-IR ISO data, however, reveal a weak OIV line, indicative
of the presence of an active nucleus (Genzel et al. 1998). The
definite presence of an AGN is seen only in X-rays, with a hard
reflected component observed below 10 keV (Iwasawa \& Comastri 1998,
hereafter IC98) and direct transmitted emission detected above this
energy (Vignati et al. 1999). The estimated unabsorbed X-ray
luminosity in the 2-10 keV energy band is $\sim 10^{44}$ erg s$^{-1}$,
implying a bolometric AGN luminosity of a few times $10^{45}$ erg
s$^{-1}$. Which of the two nuclei corresponds to the active source is
not known, and it is possible that both could harbor an AGN. This
paper provides the first direct observation in X-rays that resolves
the two nuclei and the complex extended emission.


\section{Observations}

Chandra observations of NGC6240 were obtained using the HRC-I detector
(Murray et al. 1997) on the 16th of February 2000 with an exposure
time of 8.8 ksec. The HRC-I is a microchannel plate detector optimised
for imaging but has no energy resolution. Its pixel size of $\sim
0.13\arcsec$ provides a good sampling of the telescope PSF (FWHM $\sim
0.4\arcsec$).

    \begin{figure}
    \centering
    \includegraphics[scale=0.35]{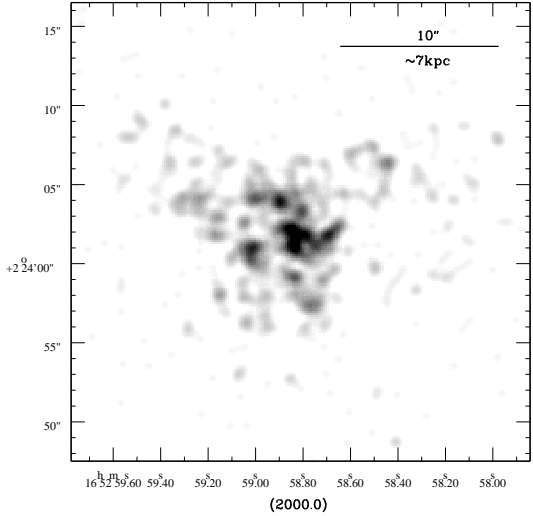}
    \caption{Smoothed Chandra HRC image of NGC6240. Regions of low
surface brightness correspond to $\protect\la 0.01$ counts
pixel$^{-1}$, while the brightest regions have $\protect\ga 0.3$
counts pixel$^{-1}$.}
    \end{figure}

HST observations of NGC6240 were obtained using the PC camera on the
23rd of March 1999 through both wide and narrow-band filters (PI: van
der Marel). Observations obtained with the F673N filter ($\Delta
\lambda = 47.2$ \AA, $\lambda_{c} = 6733$ \AA), corresponding to the
redshifted H$\alpha$ wavelength ($\lambda_{\rm H}\alpha = 6724$ \AA),
are presented here.


\section{Extended emission}

    \begin{figure}
    \centering
    \includegraphics[scale=0.35]{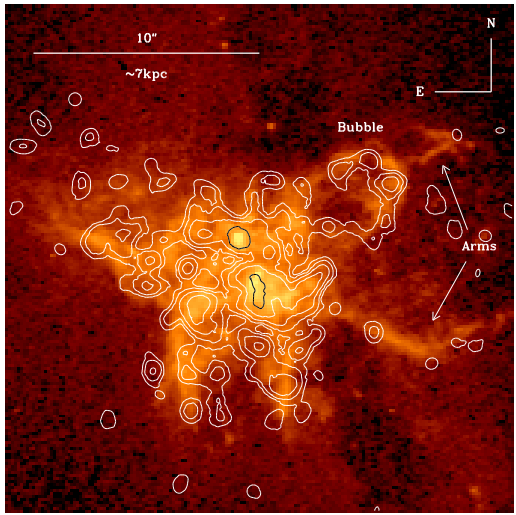}
    \caption{H$\alpha$ HST image of the central region of
NGC6240 with overlaid X-ray isocontours. The position of the
nuclei correspond to the brightest peaks of emission seen in 
the H$\alpha$ image and the associated X-ray emission is shown
using black contours.}
     \end{figure}

Figure 1 shows the central region of NGC6240 as observed by Chandra.
No other emission associated with the galaxy was detected outside this
region. A total of 509 counts were detected within a radius of
10.5\arcsec, of which about $5\%$ correspond to background emission,
which was determined from a large area free of X-ray sources.  The
data have been slightly smoothed in order to increase the
signal-to-noise using a Gaussian with $\sigma = 0.3\arcsec$. The
emission is clearly extended and presents a filamentary and clumpy
morphology with several concentrations of emission.

Figure 2 shows the HST H$\alpha$ image with X-ray contours
overlaid. The H$\alpha$ data shows remarkable structure, with very
extended filaments and loops. This complex morphology is suggestive of
gas outflowing from the distorted main plane of the galaxy, which is
roughly coincident with the orientation of the dark dust lane seen in
our figure at a PA of $\sim 15 \degr$ (east from north - for a larger
V-band image of the galaxy see Fig 1 in Colbert et al. 1994; for a
discussion on the H$\alpha$ morphology see Heckman, Armus \& Miley
1990).  Spectroscopic evidence for the presence of an outflowing
galactic wind has been discussed by Heckman, Armus \& Miley (1990) who
derived an internal velocity dispersion in excess of 1000 km s$^{-1}$,
which is highest at the position angle coincident with the minor axis
of the galaxy.

The most interesting spatial features can be seen to the west in Fig
2, where two giant `arms' can be seen in H$\alpha$ emission extending
9\arcsec\ ($\sim 6.5$ kpc) perpendicularly to the galactic plane with
an opening angle $\sim 60\deg$, suggesting a limb-brightened cone-like
outflow. The base of the northern arm presents a spectacular `bubble'
morphology with dimensions $\sim 5.2 \times 3.6\arcsec$ ($3.7 \times
2.6$ kpc). On the other hand, the eastern region has higher surface
brightness and a less clearly defined morphology, possibly resulting
from a more dense ambient medium which slows down and diverts the
outflowing wind.

The X-ray contours follow remarkably well most of the structures seen
in H$\alpha$. On the western side of the galaxy, however, most of the
X-ray emission is associated with the bubble, while the southern arm
shows little emission. This in part could be due to the presence of
the dust lane. Most of the observed X-ray emission is diffuse although
some knots with higher surface brightness are seen in the central
region (see Figure 1), particularly the areas associated with the
nuclei (see Section 4).

To obtain X-ray fluxes from the HRC observations we must assume a
spectral shape for the emission within the standard 0.08-10 keV energy
band of the instrument. From ASCA observations we know that thermal
emission and a power law are required to explain the observed X-rays
in the 0.5-10 keV energy range (IC98). The power law is very hard and
corresponds to scattered light from the obscured AGN. The thermal
emission is most probably related to the starburst activity and at
least two components are needed: a warm ($kT \sim 1$ keV) and obscured
($N_{H} \sim 10^{22}$ cm$^{-2}$) component, probably arising from the
innermost regions of the starburst, and a cool ($kT \sim 0.2-0.6$ keV)
and less absorbed ($N_{H} \sim 0.6-5 \times 10^{21}$ cm$^{-2}$)
component, associated with the more external regions of the galactic
wind (IC98). These results are in good agreement with what is
observed in other well studied starburst galaxies such as M82, Arp 299
and NGC3256 (Strickland, Ponman \& Stevens 1997, Heckman et al. 1999,
Lira et al.  2002).

The extent and morphology of the observed X-ray emission suggests that
the bulk of it is associated with an ongoing starburst. Assuming that
the thermal components observed with ASCA are a good representation
for the extended emission seen in Fig 1 (after removing the emission
associated with the nuclei), we find an absorbed flux of $\sim 1.4
\times 10^{-12}$ ergs s$^{-1}$ cm$^{-2}$ in the 0.08-10 keV energy
band (from an aperture $\sim 20\arcsec$ in diameter). The unabsorbed
flux corresponds to $\sim 0.6-1.4 \times 10^{-11}$ ergs s$^{-1}$
cm$^{-2}$ ($L_{\rm x} \sim 2-4 \times 10^{43}$ ergs s$^{-1}$),
depending on the spectral parameters of the warmer component. This
diffuse component dominates the X-ray emission in the 0.08-10 keV
energy range. The implied 0.5-2.0 keV flux corresponds to $\sim 9
\times 10^{-13}$ ergs s$^{-1}$ cm$^{-2}$, which is similar to the $6
\times 10^{-13}$ ergs s$^{-1}$ cm$^{-2}$ flux determined by IC98 in
the same spectral range.

The total H$\alpha$ flux within an aperture of 18\arcsec\ in diameter,
corresponds to $\sim 1.4 \times 10^{-12}$ ergs s$^{-1}$ cm$^{-2}$ To
correct for extinction we must estimate the value of $A_{V}$ towards
the starburst. NICMOS colour maps suggest a highly variable
extinction, with $A_{V} \sim 2-8$ (Scoville et al. 2000). This is in
agreement with the extinction map derived by Tecza et al.
(2000). Assuming an average $A_{V} = 3$, we derived a corrected total
H$\alpha$ flux of $\sim 1 \times 10^{-11}$ ergs s$^{-1}$ cm$^{-2}$,
and a total luminosity $L_{\rm H\alpha} = 3 \times 10^{43}$ ergs
s$^{-1}$.

A non-thermal radio source with flux 22.5 mJy ($\lambda = 20$cm) is
coincident with the bubble seen in the western region of NGC6240
(source W1 in Colbert et al. 1994). The source is more compact than
the structures seen in H$\alpha$ and X-rays ($\sim 1.9\arcsec$) and
from the crude positional information we extracted from Fig 3a in
Colbert et al. (1994) it appears to be located towards the top region
of the bubble. The H$\alpha$ flux associated with the bubble is $\sim
7 \times 10^{-14}$ ergs s$^{-1}$ cm$^{-2}$. Assuming $A_{V}=3$ the
corrected H$\alpha$ flux is $\sim 6 \times 10^{-13}$ ergs s$^{-1}$
cm$^{-2}$.  The observed X-ray flux is $\sim 3 \times 10^{-14}$ ergs
s$^{-1}$ cm$^{-2}$ (from an aperture $\sim 5\arcsec$ in diameter and
assuming a thermal model described by the cool component as determined by the
ASCA observations). The intrinsic X-ray flux is $\sim 5 \times
10^{-13}$ ergs s$^{-1}$ cm$^{-2}$, for $N_{H} \sim 5\times 10^{21}$
cm$^{-2}$. The H$\alpha$ to X-ray flux ratio for the bubble in NGC6240
is therefore 1.2, in good agreement with the ratios determined for
different regions of the galactic wind in NGC253, which are generally
within a factor 2 of unity (Strickland et al. 2001).



\section{The double nuclei}

    \begin{table} \caption{Nuclear fluxes in the 0.08-10 keV energy
    range} \begin{tabular}{lcc}
             \noalign{\smallskip}
             Nucleus-Model     & Observed flux          & Unabsorbed flux \\
                               & ergs s$^{-1}$ cm$^{-2}$ & ergs s$^{-1}$  cm$^{-2}$ \\
             \noalign{\smallskip}
             North-thermal           &  $1.9\times10^{-14}$ & $1.1\times10^{-13}$\\
             South-thermal           &  $8.7\times10^{-14}$ & $5.2\times10^{-13}$\\
             North-power-law         &  $1.5\times10^{-13}$ & $1.9\times10^{-13}$\\
             South-power-law         &  $7.1\times10^{-13}$ & $8.9\times10^{-13}$\\
             \noalign{\smallskip}
          \end{tabular}
    \end{table}

From the registration of the H$\alpha$ image with the HST data
obtained using the F814W filter ($\sim$ I band), which is less
affected by reddening and allows for a better determination of the
region morphology, we have identified the galaxy nuclei as the
brightest knots of emission in the H$\alpha$ image. The northern
nucleus is clearly resolved, with a FWHM $\sim 0.23\arcsec$ ($\sim
160$ pc). The southern nucleus is resolved into two knots of H$\alpha$
emission $\sim 0.44\arcsec$ apart.  Obscuration of an extended source
is probably responsible for the observed morphology in the southern
nucleus, as supported by images obtained at longer wavelengths
(Gerssen et al. 2002).

Radio maps show two bright and unresolved sources (FWHM $<0.1\arcsec$
or 70 pc) coincident to within errors with the position of the optical
nuclei, and which account for about half of the radio power emitted by
the central region of NGC6240 (Colbert et al. 1994). The highest
resolution radio observations at $\lambda = 2$cm and obtained with a
beam size of $0\farcs15\times 0\farcs14$ imply a distance of $\approx
1.5\arcsec$ ($\sim$ 1070 pc) between the two nuclear peaks (Carral et
al. 1990). This resolution is well matched by archival NICMOS
observations of NGC6240 (Scoville et al. 2000), which have a FWHM for
a point source of 0.1\arcsec-0.2\arcsec. Both nuclei appear resolved
in these images, with FWHMs $\sim 0.4\arcsec-0.5\arcsec$ ($\sim 320$
pc).  From the F160W ($\sim$ H band) and F222M ($\sim$ K band) images
($\lambda_{c} \sim 1.6$ and 2.22 $\mu$m, respectively) we find a
nuclear separation of $\approx 1.8\arcsec$ ($\sim$ 1280 pc).

The X-ray image does not have sufficient signal to noise to allow an
accurate determination of the flux-centroids or the analysis of the
spatial profile of the nuclei. In fact, only by registering the X-ray
and H$\alpha$ images it is possible to determine the position of the
X-ray nuclei. The northern nucleus is marginally resolved, with 10
counts observed within a diameter of $\approx 0.8\arcsec$ (the
instrument PSF FWHM is $\sim 0.4\arcsec$) and without showing an
obvious peak of emission in the raw data. The southern nucleus is
extended in the north-south direction and presents a complex
morphology without a clear peak but instead showing several knots of
emission. Adopting an aperture of $\approx 2.2\arcsec$ in diameter, 46
counts are detected. Background contribution is negligible within the
nuclear apertures. If the X-ray and the H$\alpha$ images are aligned
using the position of the northern nucleus (as shown in Figure 2),
then the H$\alpha$ emission from the southern nucleus seems to be
slightly displaced towards the north of the X-ray emitting region,
possibly due to absorption effects.

To convert the observed count rates to fluxes we must assume a
spectral shape for the sources. The nuclear emission could be due to
star forming activity in the innermost and most obscured part of the
starburst. In this case the emission is probably best described by the
spectral parameters of the warm and highly absorbed thermal component
as determined from the ASCA analysis (IC98). Alternatively, the nuclei
could be dominated by the AGN reflection component, which is
characterised by a hard power-law with $\Gamma =1.5$ below $\sim 3$
keV and $\Gamma =0.3$ above $\sim 3$ keV and by an absorbing column of
$N_{H} = 10^{22}$ cm$^{-2}$ (IC98). The observed and corrected fluxes
for the nuclei under these two scenarios are presented in Table 1.
Note that the value for $N_{H}$derived by IC98 might not reflect
actual physical conditions, as it depends entirely on the chosen
spectral shape. The dusty enviroment in the nuclear region of NGC6240,
on the other hand, suggests that some of this obscuration could be
real. Also, note that this column density is affecting scattered light
from the AGN, while the the active nucleus is obscured behind a gas
column of at least $10^{24}$ cm$^{-2}$.

    \begin{figure*} \centering \includegraphics[scale=0.75]{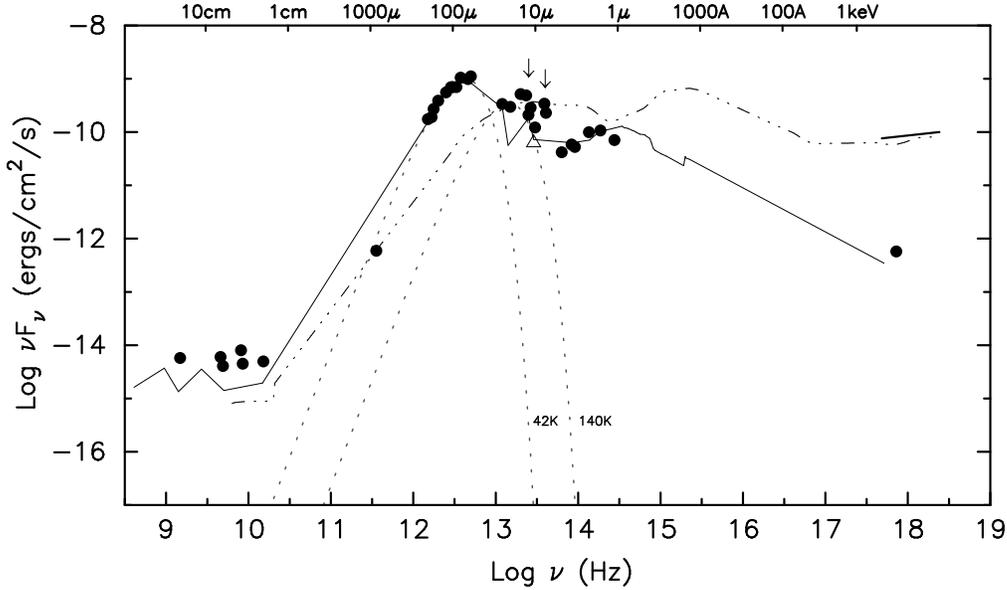}

    \caption{Observed spectral energy distribution of the NGC6240
    central region. Radio data come from Colbert et al. (1994). SCUBA
    and ISO measurements are from Klass et al. (1997) and Lisenfeld et
    al. (2000). NICMOS near-IR photometry (using the F110W, F160W and
    F222M filters) has been taken from Scoville et al. (2000). X-ray
    measurements are from this work. Nuclear N-band (10.5$\mu$m) flux
    from Krabbe et al. (2001) is shown with a white triangle.  The AGN
    unabsorbed 2-10 keV continuum has been extrapolated from the
    BeppoSAX observations (Vignati et al. 1999) and is shown with a
    thick line. A power-law index $\Gamma = 1.8$ and a flux of $4.1
    \times 10^{-2}$ photons s$^{-1}$ cm$^{-2}$ keV$^{-1}$ at 1 keV
    have been assumed for this emission. The data are compared with
    the spectral energy distribution of radio quiet quasars
    (dash-dotted line) and `high-reddening' starburst galaxies
    (continuous line) taken from Elvis et al. (1995) and Schmidt et
    al.  (1997). Two modified blackbodies with temperatures of 42 and
    140 K, and emissivity law $\propto \lambda^{-1}$, are also
    included (grey dotted lines).}

    \end{figure*}

The presence a Compton-thick AGN in NGC6240 has been determined from
ASCA and BeppoSAX observations (IC98, Vignati et al. 1999). Assuming
the spectral shape and flux observed by ASCA, about 85 counts are
predicted from the power-law scattered component in our HRC
observation. This number of counts suggests that the southern nucleus,
with 46 detected counts, is a better candidate for the counterpart of
the scattered emission. This would also be supported by HST
narrow-band observations which show that the brightest optical peak in
the southern nucleus has emission line ratios compatible with the
presence of a high excitation region (Rafanelli et al. 1997). The
discrepancy between the predicted counts and the observed numclear
values couls also be due to contamination with the large ASCA PSF. The
starburst could make a significant contribution to the power law
component if there was an unresolved component of accreting binaries,
for example. But note that the predicted HRC number of counts is very
sensitive to the assumed slope of the power-law component and its
change in value from the hard to the soft band.

A compact reflector in the southern nucleus could be associated with
the dust in a molecular torus lying as close as a few parsecs from the
central source. A large scale reflector could be associated with the
electrons in the NLR, a region hundreds of parsecs across. The size of
the X-ray region associated with the southern nucleus is not well
defined by our observations, but a reflector as large as a few
thousand parsec across is plausible. Recent Chandra observations of
the prototypical Seyfert 2 galaxy NGC1068 show hard emission arising
from a region $\sim 2$ kpc in size (Young et al. 2001). Young et
al. pointed out that results from XMM-Newton RGS observations suggest
that the gas is mainly photo-ionised, implying that its density would
be too low to provide the required scattering efficiency. It is also
possible, however, that some dense thermal gas is present, which may
be responsible for the observed hard continuum. So the scattering
scenario remains plaussible.


\section{AGN vs starburst activity}

In Figure 3 we show the {\em observed} large aperture radio, IR and
X-ray emission of NGC6240 in order to investigate its spectral energy
distribution (SED). We compare these data with the typical spectral
energy distributions of QSOs and starburst galaxies taken from Elvis
et al.  (1995) and Schmidt et al. (1997).  The starburst distribution
has been normalized to match the observed far-IR emission, while the
QSO distribution has been scaled to match the inferred 2-10 keV AGN
emission. The 10-200 $\mu$m continuum can be represented by two
modified blackbody components with emissivity $\lambda^{-1}$ with
temperatures of 42 and 140 Kelvin degrees (see Klass et al. 1997 for
details), corresponding to luminosities $L_{42} \sim 3\times10^{45}$
ergs s$^{-1}$ and $L_{140} \sim 1\times10^{45}$ ergs s$^{-1}$,
respectively (for $d=147$ Mpc). The cold component is believed to
originate from the heating of dust by OB stars in star forming
regions, while the warmer component is interpreted as emission from
the reprocessing of hard photons by dust associated with the molecular
torus located near to the active nucleus (Klass et al. 1997).

Fig 3 shows that the starburst SED is in very good agreement with the
data at most wavelengths, suggesting that the starburst component
dominates most of the observed emission. The departure of the SCUBA
data point at $\sim 1000$ $\mu$m from the model SED is not significant
as Schmidt et al. simply draw a straight line to join the radio with
the far-IR 100 $\mu$m for the average SED of their starburst
sample. The excess emission at radio wavelengths is within the large
intrinsic spread found at radio frequencies for the galaxies in the
Schmidt's sample, and also possibly contains some contribution from
the AGN, as already noted by (Colbert et al. 1994) (see also Iwasawa
et al. 2001). For example, the unresolved radio nuclei contribute
$\sim 40\%$ of the total 3.6 cm flux (Colbert et al. 1994).

A significant excess over the starburst SED is seen around the
position of the 140 K bump, consistent with the idea that this
component is related to the AGN and not to the star-forming
activity. Some contribution from the nuclear starburst, however, is
likely. PHA emission features are readily visible at wavelengths $\sim
8-12$ $\mu$m (see arrows in Fig 3; also see Fig 1 in Klaas et al. 1997
for a more detailed plot of the ISO data). Also, N-band (10.5 $\mu$m)
imaging shows that the unresolved nuclear flux (shown as a triangle)
corresponds only to 50\% of the flux measured by ISO (Krabbe et al.
2001).

In agreement with our SED, detailed observations suggest that the AGN
is completely absorbed in the near-IR, optical and soft X-ray bands
(ie, in the $\sim 2 \mu$m - 10 keV range; Tecza et al. 2000, Veilleux
et al. 1995, IC98), resulting in an observed pure starburst SED at
these wavelengths. If the absorbed AGN continuum emission is
reprocessed and emerges as the observed warm 140 K bump, the $L_{140}$
component should be of the same order as the amount of energy removed
by dust absorption at wavelengths shorter than $\sim 1$ $\mu$m. We can
test this hypothesis by integrating the QSO SED, as shown in Fig 3,
for wavelengths shorter than 1 $\mu$m and then comparing this value
with that derived for $L_{140}$. For $\lambda < 912$\AA, however, UV
photons are also absorbed by photoelectric processes. It is estimated
that about 2/3 of the ionizing continuum is absorbed by dust, either
directly or as reemited Ly$\alpha$ photons (Calzetti et al. 2000). We
find that the amount of AGN continuum removed from the QSO SED
corresponds to $\sim 4\times10^{45}$ ergs s$^{-1}$, which is 4 times
larger than the luminosity of the 140 K component.

This discrepancy between $L_{140}$ and the absorbed AGN continuum may
be even larger if, as suggested by the nuclear N-band 10.5 $\mu$m
measurements, the large ISO apertures include some significant amount
of emission from the starburst, implying an overestimation of the
$L_{140}$ emission from the nuclear region. On the other hand, the
assumption that the average QSO SED is a good representation of the
intrinsic AGN continuum distribution in NGC6240 is also highly
uncertain. In fact, the nuclear N-band measurements also imply that
the AGN component is about an order of magnitude underluminous at 10
$\mu$m if the extrapolated 2-10 keV X-ray emission (see caption to Fig
3) is used for the normalization of the QSO SED. This was already
noted by Krabbe et al. (2001), who speculated that a higher extinction
could be responsible for the faint N-band flux.  Finally, some of the
discrepancy between $L_{140}$ and the amount of removed AGN continuum
could be explained by the intrinsic properties of the AGN.  Some torus
models predict a strong anisotropy of the reprocessed emission, with
tori viewed edge-on being much fainter than tori observed pole-on, as
most of the emission is radiated from the inner face and the top and
bottom surfaces of the torus (Pier \& Krolik, 1992). This idea is
supported by recent results from an IR study of the CfA Seyfert Sample
which show clear evidence for non-isotropic emission (Perez Garcia \&
Rodriguez Espinosa, 2001). It is found that the ratio of the warm,
torus-related component, to the total IR flux is $\sim 0.4$ for
Seyfert 1 galaxies and $\sim 0.3$ for Seyfert 2 galaxies. NGC6240 has
a ratio of only 0.24 (Klaas et al. 1997), in agreement with a nearly
edge-on orientation of the dusty torus and a significant
underestimation of the intrinsic luminosity of the 140 K bump.

If our interpretation of the IR emission is correct, the luminosity of
the 42 K component should arise from the dust associated with the
starburst. Tecza et al. (2000) have determined that the starburst in
NGC6240 can be characterised by an age of $\sim 15-25$ Myr and a burst
duration $\la 5$ Myr. Using these parameters they predict a ratio of
bolometric to K-band luminosity (which has a negligible contribution
from the AGN) of $\sim 100$, and infer a bolometric luminosity for the
starburst of $\sim 6\times10^{45}$ ergs s$^{-1}$ (for $d=147$
Mpc). Since very little stellar emission is produced for wavelengths
longer than $\sim 1$ $\mu$m, this bolometric emission can be directly
compared with the observed reddened continuum at $\lambda < 1$ $\mu$m,
and the difference should correspond to the amount of stellar energy
absorbed by dust and reemited in the 42 K bump. Using the Schmidt SED
in Fig 3 as the template for the reddened spectral distribution
(parametirised as $F_{\nu} \propto \nu^{-1.8}$ for $\lambda < 1$
$\mu$m) we find that the integration under the curve gives an observed
luminosity of $\sim 6\times10^{44}$ erg s$^{-1}$, assuming, as above,
that 2/3 of the ionizing continuum ($\lambda < 912$\AA) is absorbed by
dust.  Therefore, $\sim 5.4\times10^{45}$ erg s$^{-1}$ of energy has
been removed from the intrinsic starburst emission and is then
reemited in the IR. Given all the uncertainties involved, this value
compares well with $L_{42}$.

This simple analysis of the SED of NGC6240 suggests that the starburst
and AGN components are extremely powerful, both components having
$L_{\rm bol}\sim 5\times10^{45}$ ergs s$^{-1}$. Even if the bolometric
luminosity of the active nucleus, inferred from the X-ray
observations, has been overestimated, its luminosity is probably
within a factor of a few of that of the starburst. NGC6240 contains,
therefore, a clear example of an active nucleus that plays a
significant role in the energetics of an infrared ultraluminous
galaxy, but which eludes detection at most wavelengths.




\bibliographystyle{/home/ltsun0/plt/tex/mnras/mnras}

\end{document}